# Full-deautonomisation of a lattice equation

R. Willox and T. Mase   *Graduate School of Mathematical Sciences, the University of Tokyo, 3-8-1 Komaba, Meguro-ku, 153-8914 Tokyo, Japan*

A. Ramani and B. Grammaticos   *IMNC, Université Paris VII & XI, CNRS, UMR 8165, Bât. 440, 91406 Orsay, France*

Abstract

In this letter we report on the unexpected possibility of applying the full-deautonomisation approach we recently proposed for predicting the algebraic entropy of second-order birational mappings, to discrete lattice equations. Moreover, we show, on two examples, that the full-deautonomisation technique can in fact also be successfully applied to reductions of these lattice equations to mappings with orders higher than 2. In particular, we apply this technique to a recently discovered lattice equation that has confined singularities while being nonintegrable, and we show that our approach accurately predicts this nonintegrable character. Finally, we demonstrate how our method can even be used to predict the algebraic entropy for some nonconfining higher order mappings.

1. INTRODUCTION

In [1] we introduced the method of 'full-deautonomisation' as a way of improving on the standard singularity confinement test for second-order birational mappings [2] such that the test also provides the value of their algebraic entropy [3], thereby effectively transforming it into a sufficient discrete integrability detector. This is best explained on an example.

Consider the mapping

$$x_{n+1}x_{n-1} = 1 - \frac{a}{x_n}, \quad (1)$$

for non-zero values of the parameter $a$. If it so happens that, at some iteration, $x_n$ (accidentally) takes the value $a$, the mapping yields the succession of values $\{a, 0, \infty, \infty, 0, a\}$ after which $x_{n+6}$ recovers the memory of the value of $x_{n-1}$ (and is in fact exactly equal to the latter). Because of this recovery of information, we say that the *singularity* that appeared due to the special value of $x_n$ has disappeared. We call this a *confined* singularity and associate it with integrability, which in the present case is fully justified since (1) is a well-known example of a QRT mapping [4]. However, it is easy to see that this link with integrability is tenuous at best. When we consider a similar situation for the mapping ($a \neq 0$)

$$x_{n+1}x_{n-1} = x_n - \frac{a^2}{x_n}, \quad (2)$$

we obtain the succession of values $\{\pm a, 0, \infty, \infty^2, \infty, 0, \pm a^3\}$ and $x_{n+7}$ is again regular and recovers the memory of the value of $x_{n-1}$, provided that $a^4 = 1$. (The symbol $\infty^2$ means that if we introduce a small quantity $\epsilon$ and assume that $x_n = \pm a + \epsilon$, then $x_{n+2}$ will be of the order $1/\epsilon$ and $x_{n+3}$ of the order $1/\epsilon^2$).

If $a^4 = 1$ we therefore have a confined singularity but we cannot infer integrability from it, as the mapping (2) is in fact a well-known nonintegrable one [5].

The method of full-deautonomisation consists in considering nonautonomous extensions of a mapping and in determining the precise dependence on the independent variable of the coefficients in the mapping, through the application of singularity confinement, by *requiring that the singularity pattern remain the same as for the autonomous case*. For example, assuming that the quantity $a$ in the two mappings above is a function of $n$, we find that the singularity patterns are unchanged if the following constraints are satisfied:

$$a_{n+3}a_{n-2} = a_{n+2}a_{n-1} \qquad (3)$$

for (1), and

$$a_{n+3}^2 a_{n-3}^2 = a_{n+2}^4 a_{n-2}^4 \qquad (4)$$

for (2). While the solutions of the characteristic equation for (3) are all roots of unity, this is not the case for (4). In fact, the largest root of the characteristic equation for (4) has a modulus greater than one (its exact value is $(1 + \sqrt{17} + \sqrt{2\sqrt{17} + 2})/4$, or approximately 2.08). Moreover, the logarithm of this largest root coincides exactly with the value of the algebraic entropy for this mapping, obtained in [5]. We remind the reader that the notion of algebraic entropy – which must be zero for a mapping to be integrable – was introduced based on the observation that the growth properties of the solution of a mapping are related to its integrable character [6]. If $d_n$ is the homogeneous degree of the numerator or denominator of $x_n$, the algebraic entropy of the mapping is given by the limit $\varepsilon = \lim_{n \to \infty} (\log d_n)/n$ ; the quantity $\exp(\varepsilon)$ is often referred to as the dynamical degree of the mapping.

The algebro-geometric foundation of the full-deautonomisation approach was laid in [7] and the method itself was developed in [1]. There we formulated the conjecture that if we deautonomise a second-order birational mapping with confined singularities, while preserving the singularity pattern obtained in the autonomous case and provided *sufficiently many* genuine (i.e. non gauge-removable) nonautonomous coefficients are introduced, the algebraic entropy of the mapping can be obtained from the confinement constraints on some of these coefficients. In particular, if the only characteristic roots that appear in the confinement constraints are roots of unity, then the equation should be regarded as integrable.

The expression "sufficiently many", used in this conjecture, requires some clarification. In the two examples we have given it sufficed to introduce an $n$-dependence in coefficients that are already present in the equation, in order to obtain conditions that lead to the value of the algebraic entropy. However this is not always the case. The mapping of Hietarinta-Viallet (H-V) [8] is a case in point. Its autonomous form is

$$x_{n+1} + x_{n-1} = x_n + \frac{1}{x_n^2}, \qquad (5)$$

with singularity pattern $\{0, \infty^2, \infty^2, 0\}$, and introducing an $n$-dependence in the terms of the right-hand side does not yield any interesting result. However, the only requirement when applying the full-deautonomisation approach is that the singularity pattern remain the same as in the autonomous case. A cursory examination of (5) shows that the singularity pattern $\{0, \infty^2, \infty^2, 0\}$ remains unaltered even if a constant term or a term proportional to $1/x_n$ are introduced in the mapping. With hindsight (see [1] for a more thorough account) we only analyse the effect of the latter and examine the full-deautonomisation

$$x_{n+1} + x_{n-1} = x_n + \frac{a_n}{x_n} + \frac{1}{x_n^2}, \qquad (6)$$

for which we find the confinement condition

$$a_{n+3} - 2a_{n+2} - 2a_{n+1} + a_n = 0. \qquad (7)$$

The largest root of the characteristic equation for (7) is $(3+\sqrt{5})/2$, in perfect agreement with the empirical value for the dynamical degree obtained by Hietarinta and Viallet [8] and the exact result derived by Takenawa in [9].

Having set the frame, we shall now show that the full-deautonomisation method can in fact also be applied outside the context of second-order birational mappings in which it was conceived. It has to be stressed that up to now, all algebro-geometric evidence for the aforementioned conjecture that underlies the method and, by extension, for the applicability of the full-deautonomisation method as a discrete integrability criterion, is limited to birational mappings on $\mathbb{P}^1 \times \mathbb{P}^1$ (or $\mathbb{P}^2$). In the following we shall show, on two examples, that the method can also be applied to lattice equations with confined singularities (integrable as well as nonintegrable) and, in particular, that it can also be applied to general reductions of these lattice equations to mappings of orders higher than 2.

2. Lattice equations with confined singularities

Applying singularity confinement to a lattice equation is not something new. The very phenomenon of singularity confinement was in fact discovered while analysing the singularity properties of the lattice KdV equation [2]. As a first example we shall give a succinct account of those results, geared towards the full-deautonomisation approach. Besides offering an example of an application to an integrable lattice equation, this will also allow us to explain some of the crucial technical points involved in the full-deautonomisation technique.

2.1 The case of the lattice KdV equation

Our starting point is the lattice KdV equation written directly in nonautonomous form:

$$x_{m,n} = x_{m-1,n-1} + \frac{a_{m,n-1}}{x_{m,n-1}} - \frac{b_{m-1,n}}{x_{m-1,n}}, \tag{8}$$

which we regard as an evolution equation, evolving in the northeastern direction, for inital data provided on a staircase that extends from the northwest down to the southeast in the $(m,n)$ plane, as sketched in Figure 1.

The simplest possible singularity pattern corresponds to $x$ vanishing for some pair of indices $m, n$. Taking $x_{m,n} = \epsilon$ we obtain the sequence of values

$$x_{m+1,n} \propto 1/\epsilon, \qquad x_{m,n+1} \propto 1/\epsilon, \qquad x_{m+1,n+1} \propto \epsilon,$$

and subsequent values (in the direction of the evolution) will be regular and will depend on the initial conditions provided that two conditions hold. The first one is

$$\frac{a_{m+1,n+1}}{b_{m+1,n+1}} = \frac{a_{m,n}}{b_{m,n}}, \tag{9}$$

which means that $a_{m,n} = f(m-n) b_{m,n}$ for some free function $f$. It is straightforward, however, to verify that if we introduce a gauge $x_{m,n} \to \phi(m-n) x_{m,n}$ and choose $\phi$ such that it satisfies the equation $\phi(m-n-1) = f(m-n)\phi(m-n+1)$, it is possible to ensure that $a_{m,n} = b_{m,n}$ for any $f$. In this particular gauge, the second confinement condition can then be written simply as

$$a_{m+1,n+1} - a_{m+1,n} - a_{m,n+1} + a_{m,n} = 0, \tag{10}$$

the solution of which is of course nothing but $a_{m,n} = g(m) + h(n)$, where $g$ and $h$ are two free functions. As explained in [2] the lattice KdV equation – and a lattice equation in general – has infinitely many singularity patterns. We expect however that, just as in all previously examined cases, the constraints introduced by the simplest pattern will suffice for all patterns to be confined. This was indeed verified for the first few such patterns.

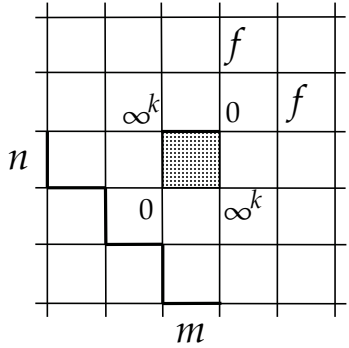

**Figure 1.** Sketch of the basic singularity pattern for the lattice KdV equation (for $k = 1$), as well as for the nonintegrable lattice equation (12) when $k \geq 2$. Here, the symbol $f$ denotes regular, non-zero values for the iterates and the bold broken line depicts a staircase pattern on which the initial values are given.

It is common knowledge that in the autonomous case, the lattice KdV equation can be put into Hirota bilinear form. That the nonautonomous lattice KdV has three different bilinearizations, the interrelations of which are not entirely clear [10,11], is perhaps not as well known. The important fact however is that (at least) one of these Hirota bilinear forms can be obtained by a straightforward reduction of a nonautonomous Hirota-Miwa (n-HM) equation (also sometimes referred to as the discrete KP equation). In [12], one of the authors (T.M.) has shown that the n-HM equation and all its reductions have the Laurent property (i.e., for a general class of initial value problems the solutions are always Laurent polynomials in the initial data). As this property extends all the way down to reductions to ordinary mappings, it is also shown in [12] that all mappings that are obtained as reductions of n-HM have at most quadratic growth and therefore have zero algebraic entropy. This result can be thought of as a discrete version of the ARS conjecture [13], at least for lattice equations associated to the Hirota-Miwa (and the discrete BKP or Miwa) equation. In particular, it proves that all reductions of the form $x_{m+p,n} = x_{m,n+q}$ of (8), for a given pair of positive integers $p, q$, have zero algebraic entropy and are therefore necessarily integrable ($p$ and $q$ are restricted to positive integers as the reduction has to be compatible with the initial value problem we consider here).

Performing the reduction $x_{m+p,n} = x_{m,n+q}$ on (8) of course also necessitates that we impose a similar condition on the parameter $a_{m,n}$ in the equation. It is then straightfoward to check that, combined with the requirement for confinement (10), this leads to the condition

$$a_{m,n+q+1} - a_{m,n+q} - a_{m,n+1} + a_{m,n} = 0, \tag{11}$$

and a similar condition in the $m$ direction, depending on $p$. Note that the case where $p = q = 1$ (which corresponds to a linearizable mapping) has to be excluded from this discussion as the singularity pattern collapses entirely and the above analysis no longer holds.

The solutions of the characteristic equation for (11) are all $q$-th roots of unity which, in the spirit of the full-deautonomisation approach, is a clear indication of the (in this case, known) integrable character of such reductions. Moreover, it is worth pointing out that regardless of the precise order of the mapping, the characteristic polynomial for (11) has a double zero at 1, which in the full-deautonomisation approach is indicative of quadratic growth for the associated reduced mapping.

This result suggests that the full-deautonomization approach might actually be applicable to lattice equations, which is quite surprising since such equations obviously contain information on the growth properties of all their reductions, almost all of which fall outside the context of second-order mappings in which the technique was first developed.

It should be noted however that the above result also shows the crucial role that gauge transformations play in our approach. Although such transformations of course have no bearing on the integrability of the lattice KdV equation, it is clear that had we not chosen the gauge $f \equiv 1$, the counterpart of condition (10) would have involved a free function of $m-n$ and no conclusion as to the integrable character of the equation could have been reached. A detailed discussion of the problems arising from gauge freedom in the case of second-order mappings and how to deal with them can be found in [14].

2.2 The case of a nonintegrable lattice equation with confined singularities

We now turn to a lattice equation recently introduced by Kanki, Mase and Tokihiro [15] which has confined singularities despite being nonintegrable, and apply our full-deautonomisation criterion. We write the equation immediately in its nonautonomous form,

$$x_{m,n} = -x_{m-1,n-1} + \frac{a_{m,n-1}}{x_{m,n-1}^k} + \frac{b_{m-1,n}}{x_{m-1,n}^k}, \tag{12}$$

where $k$ is a non-negative integer, and we consider exactly the same initial value problem as for the lattice KdV equation. For $k = 1$, this lattice equation is indeed nothing but the lattice KdV equation (up to a simple gauge transformation) and in what follows we shall not consider this integrable case and instead concentrate on the cases where $k \geq 2$, which turn out to be nonintegrable.

As in the case of KdV, the simplest singularity pattern corresponds to $x$ vanishing for some pair of indices $m, n$. Taking $x_{m,n} = \epsilon$ we obtain the sequence of values

$$x_{m+1,n} \propto 1/\epsilon^k, \qquad x_{m,n+1} \propto 1/\epsilon^k, \qquad x_{m+1,n+1} \propto \epsilon,$$

after which all subsequent values are regular and depend on the initial conditions, for all values of the exponent $k \geq 2$, provided the coefficient functions $a_{m,n}$ and $b_{m,n}$ satisfy the two very simple conditions

$$a_{m+1,n+1} = (-1)^k a_{m,n} \quad \text{and} \quad b_{m+1,n+1} = (-1)^k b_{m,n}. \tag{13}$$

Figure 1 shows this simplest pattern. We expect, just as in the case of KdV, that the constraints introduced by the simplest pattern will suffice. We illustrate this by examining the pattern shown in Figure 2, which corresponds to an initial condition where two adjacent $x$ vanish. We start by taking $x_{m+1,n} = \epsilon$ and $x_{m,n+1} = \kappa \epsilon$ and obtain the following pattern:

$$x_{m+2,n} \propto 1/\epsilon^k, \qquad x_{m+1,n+1} \propto 1/\epsilon^k, \qquad x_{m,n+2} \propto 1/\epsilon^k, \qquad x_{m+2,n+1} \propto \epsilon, \qquad x_{m+1,n+2} \propto \epsilon,$$

and again we find that all subsequent values are regular and depend on the initial conditions, provided the coefficient functions satisfy (13). The same holds true for the next few singularity patterns and we presume that all initial patterns will lead to confined singularities.

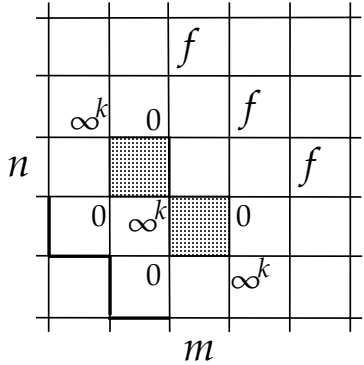

**Figure 2.** Sketch of a more complicated singularity pattern for the lattice equation (12) (for $k \geq 2$). The symbol $f$ denotes regular, non-zero values for the iterates. Note that increasing the length of the sequence of initial zeros on the diagonal only results in a similar lengthening of the entire singularity pattern.

It is clear from (13) that for even $k$ it is always possible to take $a$ and $b$ constant, whereas for odd $k$ the simplest way to satisfy the confinement conditions is to choose $a_{m,n}$ proportional to $(-1)^n$ or $(-1)^m$ (and similarly for $b_{m,n}$). We remark also that the ratio $a_{m,n}/b_{m,n}$ obeys exactly the same equation (9) as in the case of the lattice KdV equation, and thus we have $a_{m,n}/b_{m,n} = f(m-n)$ where $f$ is a free function. Here again it is possible to introduce a gauge $x_{m,n} \to \phi(m-n)x_{m,n}$ and to choose the function $\phi$ so as to have $\phi(m-n-1) = f(m-n)\phi(m-n+1)$ leading to $a_{m,n} = b_{m,n}$. In what follows we shall always work in this particular gauge.

In their paper, the authors of [15] concentrate on the autonomous case of (12) and show, based on their co-primeness criterion, that it must be nonintegrable for all even values of $k \geq 2$. It is clear that the confinement conditions (13) are not rich enough to offer us a similar insight. We therefore proceed to the full deautonomisation of (12), for arbitrary $k \geq 2$, along the same lines as in the case of the mapping (5), i.e. we allow the addition of terms to the right-hand side of (12) that do not alter the singularity patterns and consider the equation

$$x_{m,n} = -x_{m-1,n-1} + \frac{a_{m,n-1}}{x_{m,n-1}^k} + \frac{a_{m-1,n}}{x_{m-1,n}^k} + \frac{c_{m,n-1}}{x_{m,n-1}} + \frac{d_{m-1,n}}{x_{m-1,n}}. \tag{14}$$

Here, the coefficient functions $a_{m,n}$ and $b_{m,n}$ are chosen such that they satisfy the confinement conditions (13), but the functions $c_{m,n}$ and $d_{m,n}$ are still to be determined.

Implementing the singularity confinement criterion for the simplest singularity pattern (that of Figure 1) and requiring that it confines in exactly the same way as the standard deautonomisation (12), we find that $d_{m,n}$ has to be chosen such that

$$d_{m,n} = \frac{1}{k}(c_{m+1,n} + c_{m,n-1}) - c_{m+1,n-1}, \tag{15}$$

for a function $c_{m,n}$ that obeys the equation

$$c_{m+1,n+1} + 2c_{m,n} + c_{m-1,n-1} - k(c_{m+1,n} + c_{m,n-1} + c_{m-1,n} + c_{m,n+1}) = 0. \tag{16}$$

As in the case of the standard deautonomisation, more complicated initial singularities do not appear to lead to new constraints on $c_{m,n}$ or $d_{m,n}$.

## 3. REDUCTIONS AND NONINTEGRABILITY

Following the approach of full-deautonomisation as developed for ordinary difference equations, we now turn to the study of the equation for the coefficient function $c_{m,n}$. Studying the growth of the solutions of (16) is not something that is impossible. However some fundamental difficulties arise immediately. First, we do not have any rigorous results, like in the one-dimensional case, linking the growth of the solutions of the equation for the coefficients to that of the solutions of the lattice equation itself. Moreover, any growth obtained cannot be linked to the algebraic entropy of the lattice equation, since such an entropy cannot be defined in an unambiguous way. (We are aware of the work of Viallet [16] who has tried to formalise empirical results previously obtained by two of the present authors. Still, despite this interesting approach, we believe that a rigorous and useful definition of algebraic entropy for multidimensional systems is still lacking).

In order to proceed with the treatment of equation (16) we shall therefore consider reductions of the lattice equation (12) of the type $x_{m,n+\ell} = x_{m+1,n}$ where $\ell$ is an integer greater than 1. Implementing the same reduction condition on the coefficient funtion $a_{m,n}$, equation (12) is then reduced to

$$x_{n+\ell} = -x_{n-1} + \frac{a_{n+\ell-1}}{x_{n+\ell-1}^k} + \frac{a_n}{x_n^k}, \tag{17}$$

for $k \geq 2$ (where we omitted the index $m$, common to all terms). Note that the reduction corresponding to $\ell = 1$ is excluded because it would make the last two terms in equation (17) coincide, a fact which completely modifies the singularity properties of the equation and invalidates the results of the analysis in the previous section.

After reduction, the function $a_n$ in (17) obeys the relation $a_{n+\ell+1} = (-1)^k a_n$. Hence, the above mapping will have confined singularities in its autonomous form, only when $k$ is even. It should be noted also that when $\ell$ is even, this mapping can be trivially integrated to the $\ell$-th order mapping

$$\sum_{j=0}^{\ell} (-1)^j x_{n+j-1} - \sum_{j=0}^{\ell-2} (-1)^j \frac{a_{n+j}}{x_{n+j}^k} = c(-1)^n, \tag{18}$$

for some integration constant $c$.

We shall however study the mapping in its general (and simpler) form (17), of order $\ell + 1$, and we shall try to assess its integrability from its corresponding full-deautonomisation

$$x_{n+\ell} = -x_{n-1} + \frac{a_{n+\ell-1}}{x_{n+\ell-1}^k} + \frac{a_n}{x_n^k} + \frac{c_{n+\ell-1}}{x_{n+\ell-1}} + \frac{d_n}{x_n}, \tag{19}$$

where $c_n$ and $d_n$ are required to satisfy the reduced forms of conditions (15) and (16):

$$d_n = \frac{1}{k}(c_{n+\ell} + c_{n-\ell}) - c_{n+\ell-1}, \tag{20}$$

and

$$c_{n+\ell+1} + 2c_n + c_{n-\ell-1} - k(c_{n+1} + c_{n-\ell} + c_{n+\ell} + c_{n-1}) = 0. \tag{21}$$

The crucial relation that needs to be analysed is the characteristic equation for condition (21):

$$(\lambda^{\ell+1} + 1)(\lambda^{\ell+1} - k\lambda^\ell - k\lambda + 1) = 0, \tag{22}$$

and in particular its second factor (where $k, \ell \geq 2$)

$$P_\ell = \lambda^{\ell+1} - k\lambda^\ell - k\lambda + 1, \qquad (23)$$

which is trivially solved when $\ell = 2$: $P_2 = (\lambda + 1)(\lambda^2 - (k+1)\lambda + 1)$. For odd $\ell \geq 3$ it is easy to see that $P_\ell$ has exactly two real roots, $\lambda_*$ and $1/\lambda_*$, for some $\lambda_* > k \geq 2$. For even $\ell \geq 4$ it has exactly three real roots: -1, and again two reciprocal ones, $\lambda_*$ and $1/\lambda_*$, for $\lambda_* > k \geq 2$. Note that according to the full-deautonomisation approach this would mean that all these mappings have positive algebraic entropy, given by the value of $\log \lambda_*$, and will therefore be nonintegrable. However, we should stress that the conjecture (as explained in the introduction) that underlies this approach was only motivated in the case of second-order mappings, and that a priori there is no reason to assume that it will remain true for higher order ones.

For second-order autonomous mappings it is known [17] that the dynamical degree which dictates the (asymptotic) degree growth of the mapping is either 1, a Pisot number (a real algebraic integer greater than 1, the conjugates of which all lie in the open unit disc) or a Salem number (a real algebraic integer greater than 1, the conjugates of which all lie in the closed unit disc with at least one on the boundary). If the mapping is confining, then the dynamical degree is either 1, a reciprocal quadratic integer or a Salem number, and this property can in fact be shown to hold for confining nonautonomous second-order mappings as well [18]. For higher order mappings the restrictions on the dynamical degree are much less strict, but although some examples with non-Salem (or Pisot) dynamical degrees are known, Salem and Pisot numbers do seem pervasive, even for higher orders (see e.g. [19] and references therein). For the polynomial $P_\ell$ it can be shown that all its complex roots indeed lie on the unit circle, whenever $k \geq 2$ and $\ell \geq 3$. This means that its largest root $\lambda_*$ is always either a reciprocal quadratic integer or a Salem number and, in fact, as far as we were able to check it appears to be always a Salem number when $k \geq 2$ and $\ell \geq 3$.

In the remainder of this section we shall verify, on a number of examples, that the full-deautonomisation approach is indeed valid for the reduced equations (17), and this even in the context of higher order mappings. In particular, this result implies that the dynamical degree for the confining mappings (17), as predicted by (22), is either a reciprocal quadratic integer (for $\ell = 2$) or a Salem number when $\ell \geq 3$.

In the simplest case, i.e. $\ell = k = 2$ for which it suffices to take $a_n = 1$, we obtain from (17) the third order mapping

$$x_{n+2} = -x_{n-1} + \frac{1}{x_{n+1}^2} + \frac{1}{x_n^2}, \qquad (24)$$

which can actually be integrated to the H-V mapping (5). In this case, we are therefore in the presence of a second-order mapping and we find that $P_2$ as given by (23), is exactly the characteristic polynomial for condition (7), its largest root being equal to $(3 + \sqrt{5})/2$ the logarithm of which indeed coincides with the value of the algebraic entropy for the H-V mapping. As a matter of fact, the logarithm of the largest root of equation (22) can be seen to coincide exactly with the algebraic entropy for all mappings obtained from $\ell = 2$, for general even $k$ (for which the entropy was proven rigorously in [20]).

Next, we consider the $k = 2$ case, with $a_n = 1$, for the reductions corresponding to $\ell = 3, 4$ and 5. To simplify the computation of the degree growth, we started each time from $x_n, x_{n+1}, \cdots x_{n+\ell-1}$ with degree 0 and we introduced homogeneous coordinates for $x_{n+\ell}$ in the form $p/q$. For $\ell = 3$ we find the sequence of degrees: 0, 0, 0, 1, 2, 4, 10, 25, 56, 128, 296, 681, 1562, $\cdots$ and a growth ratio (which converges to the dynamical degree of the mapping and the logarithm of which therefore gives the algebraic entropy) of roughly 2.29, in agreement with the value $\lambda_* = (1 + \sqrt{3} + \sqrt{2\sqrt{3}})/2$ for the largest root of (22), obtained from $P_3$. For $\ell = 4$ we find 0, 0, 0, 0, 1, 2, 4, 8, 18, 41, 88, 188, 404, 872, $\cdots$ and a growth ratio of 2.15, again agreeing nicely with the value $\lambda_* = (3 + \sqrt{5} + \sqrt{6\sqrt{5} - 2})/4$ obtained from $P_4$. Finally, in the case $\ell = 5$ we obtain the sequence 0, 0, 0, 0, 0, 1, 2, 4, 8, 16, 34, 73, 152, 316, 656, $\cdots$ with a ratio of 2.08, in agreement with the largest root $\lambda_* = (1 + \sqrt{17} + \sqrt{2\sqrt{17} + 2})/4$ of $P_5$.

We turn now to the $k = 3$ case and examine first the $\ell = 2$ reduction. Note that since $k$ is odd, one cannot take $a_n = 1$ any more. However for $\ell = 2$ the choice $a_n = (-1)^n$ suffices. The corresponding third-order mapping can be integrated to one of the H-V extensions introduced in [1], the algebraic entropy of which is equal to $\log(2 + \sqrt{3})$. This coincides exactly with the result we obtain from (22). Next, we consider the case $\ell = 3$. Here $a_n$ obeys the relation $a_{n+4} = -a_n$ and should be chosen accordingly, for instance by choosing for 4 consecutive $a$'s the value $+1$, for the next 4 the value $-1$ and repeating the pattern. We obtain thus the sequence of homogeneous degrees: 0, 0, 0, 1, 3, 9, 30, 100, 324, 1053, 3429, $\cdots$ with a ratio of 3.256 in agreement with the value of $(3 + \sqrt{17} + \sqrt{10 + 6\sqrt{17}})/4 \sim 3.254$ obtained from $P_3$ for $k = 3$.

At this point a remark is in order. One could wonder what would have happened had we taken an $a_n$ that violates the constraint $a_{n+4} = -a_n$, for instance by choosing $a_n = 1$ throughout. The singularity analysis is a useful guide in this case. Starting for regular $x_n$, $x_{n+1}$, $x_{n+2}$ and $x_{n+3} = 0$, we obtain, when $a_n$ obeys the correct relation, the confined singularity pattern $\{0, \infty^3, f, \infty^3, 0\}$ where $f$ is a finite value. If however $a_n$ is chosen incorrectly, we find a nonconfined singularity pattern of the form $\{0, \infty^3, f, \infty^3, 0, \infty^3, f', \infty^3, 0, \infty^3 \cdots\}$. Given the length of the confined singularity pattern we expect the first difference in the degree, compared to that obtained for the correct choice of $a_n$, to occur at the level of the degree 324, and in fact when we perform the calculation we obtain a value of 327, followed by degrees 1071, 3513, ... which corresponds to a growth ratio of 3.28. Using the method introduced in [14] in relation to the late confinement property, we can actually establish an equation that gives the degree growth, even in this nonconfining case. The way to proceed is to view the nonconfining pattern as the limit of a sequence of patterns obtained by postponing confinement by an extra four steps each time, i.e. $\{0, \infty^3, f, \infty^3, 0, \infty^3, f', \infty^3, 0\}$, $\{0, \infty^3, f, \infty^3, 0, \infty^3, f', \infty^3, 0, \infty^3, f'', \infty^3, 0\}$, etc. Skipping the detail of the singularity confinement analysis, the characteristic polynomials one obtains from the conditions on the coefficients are all of the form $(\lambda^4 + 1)P_m(\lambda)$ for

$$P_m(\lambda) = 1 + \sum_{j=1}^{m} \lambda^{4j-3}(\lambda^3 - 3\lambda^2 - 3) = 1 + \lambda\frac{\lambda^{4m} - 1}{\lambda^4 - 1}(\lambda^3 - 3\lambda^2 - 3), \qquad (25)$$

where the case $m = 1$ corresponds to the basic confinement pattern that gave rise to (22), $m = 2$ to the pattern that is 4 steps longer, and so on. It is easy to see that all these polynomials have at least one root that is greater than 1 and that the value of their largest root increases with increasing $m$ and, in fact, tends to the largest root of $\lambda^3 - 3\lambda^2 - 3 = 0$. The value obtained by solving this equation numerically is 3.279, in excellent agreement with the value obtained by direct calculation of the degrees (and, it should be noted, significantly different from the value 3.254 obtained for the basic confining pattern). It turns out that the dynamical degree we obtain for the nonconfining case is in fact a Pisot number. This, together with the fact that the dynamical degree for the confining mapping is a Salem number, of course leaves open the possibility that (17) with $k = 3 = \ell$ might not be a primitive 4th order mapping, and that in some special cases it might be possible to integrate it (twice) to a second-order mapping.

Let us give two more interesting examples, in order to strengthen our arguments. First we examine the case $k = 3$ with $\ell = 4$. Taking $a_n = (-1)^n$ we find the degree sequence 0, 0, 0, 0, 1, 3, 9, 27, 84, 262, 810, 2502, $\cdots$ with a ratio of 3.09, in excellent agreement with the exact value for the larget root of (22): $\lambda_* = (2 + \sqrt{2} + \sqrt{4\sqrt{2} + 2})/2 \sim 3.0907$. Here as well we can predict the dynamical degree for the nonconfining case of the mapping: it is the largest root of the polynomial $\lambda^4 - 3\lambda^3 - 3$, which is approximately equal to 3.1006 and *not* a Pisot number. The value we predict from our approach fits perfectly with the degree growth observed by iterating the mapping (with $a_n = 1$) 0, 0, 0, 0, 1, 3, 9, 27, 84, 262, 813, 2520, $\cdots$, which yields a growth ratio of 3.10. The fact that the predicted dynamical degree is not a Pisot number strongly suggests that this mapping is indeed of higher order and that our approach therefore enables us to predict the algebraic entropy for higher order mappings as well.

The last case we treat, $k = 3$ with $\ell = 5$, is confining under the constraint $a_{n+6} = -a_n$. It leads to the degree sequence 0, 0, 0, 0, 0, 1, 3, 9, 27, 81, 246, 748, 2268, 6876, $\cdots$ with a ratio of 3.032, again in very good agreement with the approximate numerical value of 3.0316 for the largest root of equation (22). Again, the predicted dynamical degree (approximately 3.0353) for the nonconfining case (which is the largest root of $\lambda^5 - 3\lambda^4 - 3$ and which fits nicely with the growth ratio of $6894/2271 \sim 3.036$ obtained after 8 iterations with $a_n = 1$) is not a Pisot number and this mapping is therefore, most probably, not reducible to a second-order mapping either.

Most importantly however, since all the reductions we analysed above turn out to be nonintegrable, we believe that in the spirit of the ARS conjecture, it is justified to conclude that the lattice equation (12), for general $k$, is indeed nonintegrable. This corroborates the result obtained in [15] for the autonomous case when $k$ is even.

## 4. Conclusions

In this paper we have examined a lattice equation, recently introduced by Kanki, Mase and Tokihiro, which has the interesting property of possessing only confined singularities, while being nonintegrable. In that paper the claim for nonintegrability of the lattice equation was established on the basis of the co-primeness methods developed by those same authors. Here we have analysed the equation using the method of full-deautonomisation which we recently proposed as a reliable discrete integrability criterion. The interesting conclusion is that our results on the integrable character of the various reductions of this lattice equation are in perfect agreement with those of the authors of [15]. In particular equation (22), which allows the computation of the algebraic entropy for the various reductions of the lattice equation, coincides with an equation that the authors of [15] have derived using the co-primeness arsenal for the autonomous case. In our mind, our results coupled to those of [15] establish beyond any doubt the nonintegrability of the lattice equation (12).

While studying the various reductions of (12) we were led to apply the full-deautonomisation method to mappings of order higher than two. Despite the fact that a rigorous justification for our approach has been presented only in the case of second-order mappings, it turns out that the method works perfectly also for mappings of higher order, whether they be integrable or nonintegrable. This greatly reinforces the usefulness of this approach as a heuristic discrete integrability detector. Moreover, we were also able to accurately predict the algebraic entropy for certain nonconfining mappings that can be thought of as limiting cases of confining ones, and this even in the case of higher order mappings.

On the other hand, while the full-deautonomisation of the lattice equation could be performed in a straightforward way, we were constrained to introduce reductions. It would be interesting, although this definitely constitutes a tall order, to be able to obtain indications of nonintegrability directly from the linear (lattice) equation for the coefficients in the fully deautonomised mapping, without having to resort to reductions. We hope to be able to address this challenge is some future work of ours.

Acknowledgements

RW would like to acknowledge support from the Japan Society for the Promotion of Science (JSPS), through the the JSPS grant: KAKENHI grant number 15K04893. TM would also like to acknowledge support from JSPS through the Grant-in-Aid for Scientific Research 25-3088.

References

[1] A. Ramani, B. Grammaticos, R. Willox, T. Mase and M. Kanki, J. Phys. A 48 (2015) 11FT02.

[2] B. Grammaticos, A. Ramani and V. Papageorgiou, Phys. Rev. Lett. 67 (1991) 1825.

[3] M. Bellon and C-M. Viallet, Comm. Math. Phys. 204 (1999) 425.

[4] G.R.W. Quispel, J.A.G. Roberts and C.J. Thompson, Physica D34 (1989) 183.

[5] C-M.Viallet, Int. J. Geom. Methods Mod. Phys. 5 (2008) 1373.

[6] A.P. Veselov, Comm. Math. Phys. 145 (1992) 181.

[7] T. Mase, R. Willox, B. Grammaticos and A. Ramani, Proc. Roy. Soc. A 471 (2015) 20140956.

[8] J. Hietarinta and C-M. Viallet, Phys. Rev. Lett. 81 (1998) 325.

[9] T. Takenawa, J. Phys. A 34 (2001) 10533.

[10] K. Kajiwara and Y. Ohta, J. Phys. Soc. Jpn. 77 (2008) 054004.

[11] K. Kajiwara and Y. Ohta, RIMS Kôkyûroku Bessatsu B13 (2009) 53.

[12] T. Mase, J. Math. Phys. 57 (2016) 022703.

[13] M.J. Ablowitz, A. Ramani and H. Segur, Lett. Nuov. Cim. 23 (1978) 333.

[14] B. Grammaticos, A. Ramani, R. Willox, T. Mase and J. Satsuma, Physica D 313 (2015) 11.

[15] M. Kanki, T. Mase and T. Tokihiro, *Singularity confinement and chaos in two-dimensional discrete systems* (2015) arXiv:1512.09168 [nlin.SI].

[16] C-M.Viallet, *Algebraic entropy for lattice equations* (2006) arXiv:math-ph/0609043.

[17] J. Diller and C. Favre, Am. Journal of Math. 123 (2001) 1135.

[18] T. Mase, *Studies on spaces of initial conditions for nonautonomous mappings of the plane and singularity confinement*, Doctoral thesis (2016) Graduate School of Mathematical Sciences, the University of Tokyo.

[19] K. Oguiso and T.T. Truong, J. Math. Sci. Univ. Tokyo 22 (2015) 361.

[20] M. Kanki, T. Mase and T. Tokihiro, J. Phys. A 48 (2015) 355202.